\documentclass[10pt,journal,final,twoside]{IEEEtran}

\IEEEoverridecommandlockouts
\usepackage{mysymbol}
\usepackage[T1]{fontenc}
\usepackage{textcomp,times}
\usepackage{amsmath,amssymb,amsfonts,amsthm,mathtools,bm,breqn}
\usepackage{graphicx,epstopdf,caption,subfig}
\usepackage{algorithm,algorithmic}
\usepackage{url,xspace,fancybox,verbatim,comment,mwe,soul}
\usepackage{array,multirow,makecell}
\usepackage{paralist,mdwlist,lineno}
\usepackage{cite}
\usepackage{empheq}
\usepackage{eurosym}
\usepackage[super]{nth}
\usepackage{xcolor}
\usepackage[colorlinks=true,
            linkcolor=blue,
            urlcolor=blue,
            citecolor=red,
            anchorcolor=blue]{hyperref}
            
\DeclareGraphicsExtensions{.pdf,.png,.jpg}
\graphicspath{{./Figures/}}

\newtheorem{remark}[theorem]{Remark}

\newcommand\upj{\mathord{\mathrm{j}}}
\allowdisplaybreaks

\begin{document}
\title{Sparse Neural Approximations for Bilevel \\ Adversarial Problems in Power Grids
}
\author{\IEEEauthorblockN{Young-ho Cho},~\IEEEmembership{Student Member, IEEE}, \IEEEauthorblockN{Harsha Nagarajan},~\IEEEmembership{Senior Member, IEEE},\\
\IEEEauthorblockN{Deepjyoti Deka},~\IEEEmembership{Senior Member, IEEE}, and
\IEEEauthorblockN{Hao Zhu},~\IEEEmembership{Senior Member, IEEE}

\thanks{\protect\rule{0pt}{3mm} The authors gratefully acknowledge support from the U.S. Department of Energy's Laboratory Directed Research \& Development (LDRD) program at Los Alamos National Laboratory under project number 20230091ER, titled ``Learning to Accelerate Global Solutions for Non-Convex Optimization'' and U.S. National Science Foundation under Awards 2130706 and 2150571.}

\thanks{\protect\rule{0pt}{3mm} Y. Cho and H. Zhu are with the Chandra Family Department of Electrical \& Computer Engineering, The University of Texas at Austin, TX, 78712, USA; Emails: {\{\href{mailto:jacobcho@utexas.edu}{jacobcho}, \href{mailto:haozhu@utexas.edu}{haozhu}\}{@}utexas.edu}.}
\thanks{\protect\rule{0pt}{3mm} H. Nagarajan is with the Applied Mathematics and Plasma Physics (T-5) group, Los Alamos National Laboratory, Los Alamos, NM, USA. Email:  \href{mailto:harsha@lanl.gov}{harsha@lanl.gov}.}
\thanks{\protect\rule{0pt}{3mm} D. Deka is with MIT Energy Initiative, MIT, Cambridge, MA, USA. Email:  \href{mailto:deepj87@mit.edu}{deepj87@mit.edu}.}
}
\maketitle

\begin{abstract}

The adversarial worst-case load shedding (AWLS) problem is pivotal for identifying critical contingencies under line outages. It is naturally cast as a bilevel program: the upper level simulates an attacker determining worst-case line failures, and the lower level corresponds to the defender’s generator redispatch operations. Conventional techniques using optimality conditions render the bilevel, mixed-integer formulation computationally prohibitive due to the combinatorial number of topologies and the nonconvexity of AC power flow constraints. To address these challenges, we develop a novel single-level optimal value-function (OVF) reformulation and further leverage a data-driven neural network (NN) surrogate of the follower’s optimal value. To ensure physical realizability, we embed the trained surrogate in a physics-constrained NN (PCNN) formulation that couples the OVF inequality with (relaxed) AC feasibility, yielding a mixed-integer convex model amenable to off-the-shelf solvers. To achieve scalability, we learn a sparse, area-partitioned NN via spectral clustering; the resulting block-sparse architecture scales essentially linearly with system size while preserving accuracy. Notably, our approach produces near-optimal worst-case failures and generalizes across loading conditions and unseen topologies, enabling rapid online recomputation. Numerical experiments on the IEEE 14- and 118-bus systems demonstrate the method’s scalability and solution quality for large-scale contingency analysis, with an average optimality gap of 5.8\% compared to conventional methods, while maintaining computation times under one minute.

\end{abstract}

\begin{IEEEkeywords}
Adversarial worst-case load shed, bilevel optimization, physics-constrained sparse neural networks, value-function approximation.
\end{IEEEkeywords}

\section{Introduction}
\IEEEPARstart{C}{ontingency} analysis is fundamental for strengthening grid resilience, especially against rising threats like natural disasters \cite{wang2015research} and cyber-physical attacks \cite{liu2015trilevel}. Identifying worst-case line contingencies is essential to prioritize maintenance and safeguard critical infrastructure from catastrophic shutdowns. Of particular interest is the adversarial worst-case load shed (AWLS) problem \cite{aquino2023comparing} that identifies the set of line contingencies of given cardinality that induces the largest load shed on the grid. Rapidly identifying such contingencies is increasingly vital for real-time operational decisions, including short-term maintenance prioritization and security measures during operational emergencies \cite{zhou2023most}.  

The AWLS problem is typically formulated as a challenging adversarial bilevel optimization, where the \emph{upper-level (leader)} problem represents an adversary disconnecting lines to maximize load shed, while the \emph{lower-level (follower)} problem minimizes load shed via optimal redispatch under power flow constraints. Formally, this problem is defined by upper-level adversarial topology variables $\bbx \in \{0,1\}$ constrained within the budget set $\mathcal{X}_k$, and lower-level operating variables $\bby$ that are constrained by the AC feasibility set $\bby\in \mathcal{Y}(\bbx)$ for the given topology $\bbx$. The bilevel problem is then concisely described by $\max_{\bbx \in \mathcal{X}_k} \eta(\bbx)$, where $\eta(\bbx) = \min_{\bby \in \mathcal{Y}(\bbx)} \text{[load shed(\bby, \bbx)]}$, is the value-function of the lower level.

Not surprisingly, the interplay between discrete adversarial actions and nonconvex AC power flow constraints results in computationally prohibitive feasible solution spaces \cite{sinha2017review}. The brute-force approach \cite{liu2013optimal} involves enumerating all possible topologies and reducing the problem to a single-level formulation becomes computationally impractical as the system size or the adversarial action set is enlarged.
To the best of our knowledge, directly addressing the AWLS problem using the full AC power flow model remains largely unexplored in existing literature, primarily due to high computational complexity. Most AC‑based studies therefore restrict either the network size or the adversarial action set to regain tractability.
For example, fast screening of contingencies via sensitivity‑based pruning can dramatically reduce candidate pairs~\cite{kaplunovich2016fast}, and sampling only the most severe single‑line outages yields up to 100× acceleration for N–1 contingency screening on small‑scale systems~\cite{du2019achieving}.

To address these challenges, recent work has pursued efficient approximations for the adversarial problem. A common step is to replace the nonconvex AC power flow with the linear DC approximation in the lower level to improve scalability. However, even under the DC model the resulting problem remains NP-hard, so the computational intractability persists despite linearization, and the approximation can also degrade accuracy \cite{johnson2022scalable,bienstock2010nk}.
Instead of linear approximations, topology-aware data-driven surrogates, such as deep neural networks (NNs), can model grid physics and alleviate the computational intractability arising from adversarial line failures \cite{cho2023topology,jalving2024physics, pareek2023fast}.
To further reduce complexity, some studies restrict the adversary’s action space to the most critical $k$-element failures. For example, focusing on a reduced search space by screening $k$-line contingencies can retain the worst-case load-shed scenario while reducing the search space to about $0.14\%$ of all four-line combinations on the IEEE 118-bus system~\cite{sarkar2022analytical}.

More recently, an NN surrogate, $NN(\bbx)$, was trained to approximate the value-function $\eta(\bbx)$ from data generated under diverse topological attacks with AC power-flow constraints \cite{aquino2023comparing}. Beyond the AWLS problem, this value-function surrogate approach is well established both in other power-grid applications \cite{vlah2023solving} and in generic bilevel optimization \cite{molan2023using}. Because ReLU networks are piecewise linear, the trained model admits a mixed-integer linear programming (MILP) reformulation \cite{grimstad2019relu}. Consequently, by maximizing the output of $NN(\bbx)$ over $\bbx \in \mathcal{X}_k$, one obtains an approximate solution to the bilevel problem. This surrogate-based approach faces two key limitations on large systems. First, prediction error is operating-point (e.g. varying loads) dependent and, under distribution shift to unseen contingencies, the surrogate can misrank topologies; as a result, directly maximizing $NN(\bbx)$ may select attack sets whose predicted load shed substantially over/under estimates the true $\eta(\bbx)$, yielding inaccurate worst-case certificates. Second, when one embeds the surrogate exactly as an MILP, the number of binary variables scales with the number of ReLU neurons, so networks sized for large grids lead to rapidly growing MILPs that become computationally burdensome. These challenges motivate formulations that couple the surrogate with physics-based feasibility constraints and adopt NN architectures that control MILP size without sacrificing performance guarantees.

To this end, we develop a scalable yet accurate data-driven approach to the AWLS problem that centers on a single-level value-function reformulation of the bilevel program. 
In our reformulation, the optimization over topologies is coupled with (i) the AC power-flow feasibility set $\mathcal{Y}(\bbx)$ (or a tight convex relaxation thereof) and (ii) a learned NN surrogate of the lower level's optimal value $\eta(\bbx)$. This coupling makes the surrogate a physics-constrained NN (PCNN): any candidate topology must admit an operating point satisfying grid physics, while the realized load shed is tethered to the value-function surrogate through explicit constraints. This PCNN-based reformulation results in improved ranking of worst-case topologies across varying loads, and better generalization to unseen contingencies.

To ensure scalability, we design a sparsified spectral clustering-based NN for $\eta(\bbx)$ approximation. Using spectral clustering to partition the grid, we zero out cross-area connections and add only light tie-line features, so the number of trainable parameters—and, after ReLU encoding, the number of MILP binaries—grows essentially linearly with system size~\cite{sanchez2014hierarchical,cho2024data}.
This design is motivated by dominant electrical locality: while line outages can have non-local impacts, their strongest redistribution effects often occur on electrically nearby branches; tie-line status and electrical-parameter features are retained to preserve the main inter-area couplings.
This modular surrogate keeps the mixed-integer reformulations tractable on large networks. Together, the physics-constrained value-function reformulation and the sparse NN architecture yield single-level formulations that are both scalable and accurate across the testing set.

We validate the improved accuracy of our NN models for value-function approximation, and for solving the AWLS problem, on the IEEE 14- and 118-bus test cases~\cite{IEEE_case_ref}. We also demonstrate robustness in a challenging setting where the adversarial action set includes new and larger topological decisions not present in training. Across all test systems, the proposed method reduces optimality gaps to below 20\%, about half those obtained by directly maximizing a standalone NN surrogate, while keeping solve times under one minute.

In a nutshell, our main contributions are three-fold:
\begin{enumerate}
    \item We learn a novel physics-constrained NN (PCNN) surrogate for $\eta(\bbx)$ and embed it via a value-function reformulation together with AC feasibility via convex relaxations, yielding a single-level mixed-integer convex program that provides accurate, physics-constrained worst-case topologies.
    \item We design a spectral clustering-based sparse NN to reduce the model complexity by exploiting the weak load coupling among areas, that enables our approach to scale up to large networks.
    \item We validate the performance of our NN model in two IEEE test systems and show that compared to existing efforts, it extrapolates favorably to unseen and larger adversarial action sets in test data. This highlights its application as a practical approach for large-scale grid topology optimization tasks.  
\end{enumerate}

The paper is organized as follows: Section~\ref{sec:bilevel} introduces the AWLS formulation. Section~\ref{sec:NN} discusses the NN-based value-function approximation and its MILP form. Section~\ref{sec:approx} develops our two novel contributions: the physics-constrained NN (PCNN) surrogate formulation and the sparse NN architecture using area partitioning. The efficacy of these methods is demonstrated in Section~\ref{sec:results} through numerical results. Finally, Section~\ref{sec:con} concludes the paper.

\section{Problem Formulation} 
\label{sec:bilevel}
Consider a transmission system consisting of $N$ buses collected in the set $\mathcal{N} := \{1,\dots, N\}$ and $L$ branches in $\mathcal{L} := \{(i, j)\} \subset \mathcal{N} \times \mathcal{N}$, where each physical line is modeled by a single directed arc $(i,j)$.
For each bus $i\in\mathcal{N}$, let $V_i\angle\theta_i$ denote the complex nodal voltage phasor, and  $\{P_i,Q_i\}$ denote the active and reactive power injections, respectively.
The power injections are calculated by the differences between the active and reactive power generations $\{P^G_i, Q^G_i\}$ and loads $\{P^D_i, Q^D_i\}$ that are denoted as $P_i:=P^G_i-P^D_i$ and $Q_i:=Q^G_i-Q^D_i$.
For each branch $(i,j)\in\mathcal{L}$, let $x_{ij}\in \{0,1\}$ denote the branch status (0/1: off/on) and $\theta_{ij}:= \theta_i-\theta_j$ denote the angle difference between buses $i$ and $j$, with $\{P_{ij},Q_{ij}\}$  the active and reactive power flows from bus $i$ to $j$.
In addition, the branch's series and shunt admittance values are respectively denoted by $y_{ij}= g_{ij}+\upj b_{ij}$ and $y_{ij}^{\mathrm {sh}}=g^{\mathrm {sh}}_i+\upj b^{\mathrm {sh}}_i$.

The power flows of branch $(i,j)$ depend on $\{V_i, V_j,\theta_{ij}\}$ with the corresponding branch status $x_{ij}$, and in the case of a transformer, its tap ratio $a_{ij}$, as given by
\begin{subequations}
\small
\label{eq:powerflow}
\begin{align}
    P_{ij}\! &=\! x_{ij}\!\left(V^2_{i} \Big(\frac{ g_{ij}}{ a^2_{ij}}\!+\!g^{\mathrm {sh}}_i\Big)\!-\! \frac{ V_{i}  V_{j}}{ a_{ij}} \big( g_{ij}\cos \theta_{ij}\!+\!b_{ij}\sin \theta_{ij}\big)\right)\!,\\
    Q_{ij}\! &=\! x_{ij}\!\left(\!- V^2_{i} \Big(\frac{ b_{ij}}{a^2_{ij}}\!+\!b^{\mathrm {sh}}_i\Big) \! -\! \frac{ V_{i}  V_{j}}{ a_{ij}} \big( g_{ij}\sin  \theta_{ij} \!- \!b_{ij}\cos  \theta_{ij}\big)\right)\!.
\end{align}
\end{subequations}
For transmission lines,  $a_{ij}$ is simply set to 1. For transformers, the tap ratio $a_{ij}$ is typically within the range of $[0.9, 1.1]$ and only affects the primary-to-secondary direction.  Thus, for the secondary-to-primary direction, one can also use $a_{ij}=1$ in \eqref{eq:powerflow}. While this work assumes $a_{ij}$ is given, it is possible to include controllable tap ratios as optimization's decision variables; see e.g.,~\cite{bazrafshan2018optimal}.

With load-shedding variables $\{\delta_i^P,\delta_i^Q\}$, the effective served demand at bus $i$ is $P_i^D-\delta_i^P$ and $Q_i^D-\delta_i^Q$. Generation re-dispatch and load shedding are modeled as corrective recourse actions: the net injection variables $P_i$ and $Q_i$ are post-contingency lower-level decision variables, while $\delta_i^P$ and $\delta_i^Q$ adjust the served demand. The AC power-flow equations and engineering limits are enforced for the surviving network. Given that $\{P_{ij}, Q_{ij}\}$ denote flows from $i$ to $j$, the nodal power balances, in terms of net injections $P_i:=P_i^G-P_i^D$ and $Q_i:=Q_i^G-Q_i^D$, are
\begin{subequations}\label{powerinj_alt}
\begin{align}
P_i + \delta_i^P &= \textstyle \sum_{(i,j)\in\mathcal L} P_{ij}, \qquad \forall i\in\mathcal N,\\
Q_i + \delta_i^Q &= \textstyle \sum_{(i,j)\in\mathcal L} Q_{ij}, \qquad \forall i\in\mathcal N.
\end{align}
\end{subequations}
Here, $P_i$ and $Q_i$ capture corrective generation re-dispatch within the generator bounds. The engineering limits, including bounds on load shedding, phase-angle differences, and an apparent-power (thermal) limit on branches, are
\begin{subequations}
\begin{align}
& \underline{V}_i \leqslant V_i \leqslant \bar{V}_i, \ \forall i\in\mathcal N,\\
& \underline{P}_i^G - P_i^D \leqslant P_i \leqslant \bar{P}_i^G - P_i^D, \ \forall i\in\mathcal N, \label{eq:net_gen_limit_p}\\ 
& \underline{Q}_i^G - Q_i^D \leqslant Q_i \leqslant \bar{Q}_i^G - Q_i^D, \ \forall i\in\mathcal N, \label{eq:net_gen_limit_q}\\
& 0 \leqslant \delta_i^P \leqslant P_i^D,\quad 0 \leqslant \delta_i^Q \leqslant Q_i^D, \  \forall i\in\mathcal N,\\
& \underline{\theta}_{ij} \leqslant \theta_{ij} \leqslant \bar{\theta}_{ij}, \quad P_{ij}^2 + Q_{ij}^2 \leqslant \bar{S}_{ij}^2, \ \forall (i,j)\in\mathcal L.
\end{align}
\label{eq:eng_limits}
\end{subequations}

\subsection{Adversarial Worst-case Load Shed (AWLS) problem:}
\label{subsec:awls_prob}
The AWLS problem identifies a topology $\bbx$ that maximizes load shedding following adversarial line failures or natural disasters. Formally, this constitutes a bilevel optimization framework where the interaction between attacker and defender is captured through hierarchical decision-making. We present this structure by first defining the adversarial objective and subsequently the system operator's response through the value-function. \\ 

\noindent
\textbf{Upper-level adversarial problem}: To maximize disruption, the attacker selects up to $k$ line failures to
\begin{align}\label{eq:awls_upper}
    \max_{\bbx \in \mathcal{X}(k)} \quad & \eta(\bbx), \quad \text{where}
\end{align}
$$\mathcal{X}(k) := \left\{ \bbx \in \{0,1\}^{|\mathcal{L}|} : \sum_{(i,j) \in \mathcal{L}} x_{ij} \geqslant |\mathcal{L}|-k \right\}.$$

\noindent
\textbf{Lower-level load-shedding problem:}
For any fixed topology $\bbx$, the value-function $\eta(\bbx)$ is the optimal total load shed attained by the operator:
\begin{subequations}
\label{eq:awls_lower}
\begin{align}
    \eta(\bbx) = \min_{\bby, \bm{\delta}^P, \bm{\delta}^Q} & \sum_{i \in \mathcal{N}} \left( \delta_i^P + \delta_i^Q \right) \label{eq:inner_obj} \\
    \text{s.t.} \quad & f(\bbx, \bby) \leqslant 0, \label{eq:inner_ineq} \\
    & h(\bbx, \bby, \bm{\delta}^P, \bm{\delta}^Q) = 0, \label{eq:inner_eq}
\end{align}
\end{subequations}

Here, $\bby$ collects the power-flow variables $\{V_i\}_{i\in\mathcal N}$, $\{\theta_{ij}\}_{(i,j)\in\mathcal L}$, $\{P_{ij},Q_{ij}\}_{(i,j)\in\mathcal L}$, and $\{P_i,Q_i\}_{i\in\mathcal N}$; $\{\delta_i^p,\delta_i^q\}_{i\in\mathcal N}$ are the active/reactive load shedding variable vectors. The constraint set in \eqref{eq:inner_ineq} enforces operational limits (voltage magnitudes, angle differences, line-flow and apparent-power bounds) with \eqref{eq:eng_limits}, while equality constraints in \eqref{eq:inner_eq} imposes the AC power-flow equations and nodal balances with \eqref{eq:powerflow}–\eqref{powerinj_alt}.

\begin{remark}
Assume full load shedding is permitted, generation re-dispatch is allowed within the bounds in \eqref{eq:net_gen_limit_p}--\eqref{eq:net_gen_limit_q}, and generator lower bounds allow zero active power output. Further assume that the reactive-power and voltage bounds admit at least one admissible operating point that balances net shunt requirements.
Under these assumptions, an admissible combination of load shedding and generation re-dispatch provides a feasible corrective operating point for every connected topology $\bbx\in\mathcal X(k)$. Then, the lower-level nonlinear program (NLP) \eqref{eq:awls_lower} is feasible for every such topology.
Consequently, $\eta(\bbx)$ is well defined and finite for all $\bbx \in \mathcal{X}(k)$, satisfying $0 \leqslant \eta(\bbx) \leqslant \sum_{i\in\mathcal N}\bigl(P_i^D+Q_i^D\bigr)$.
Since $\mathcal X(k)$ is finite, the upper-level problem \eqref{eq:awls_upper} is bounded and attains an optimal solution.
\label{rem:feasibility}
\end{remark}

\begin{remark}
In this work, we restrict the adversarial set to connected post-contingency topologies; outage combinations that create islands are therefore excluded from data sets and benchmark construction. In utility energy management systems, such events are typically identified through the topology-processing module and treated as a separate operating condition. Their accurate evaluation requires an island-aware model with island-specific power balance, voltage-angle references, and treatment of islands with insufficient active-power, reactive-power, or voltage support. We instead focus on severe load shedding in connected networks, where the response is governed by nonlinear AC feasibility, corrective redispatch, voltage constraints, and thermal limits. Extending the proposed framework and training data to island-aware AWLS is left for future work.

\label{rem:islanding}
\end{remark}

The bilevel program \eqref{eq:awls_upper}–\eqref{eq:awls_lower} combines combinatorial interdiction decisions with nonlinear, non-convex AC physics, yielding a non-convex and large-scale optimization problem. Heuristic decompositions (e.g., Benders-type \cite{salmeron2009worst}) may lack global guarantees on realistic networks, whereas Karush-Kuhn-Tucker (KKT) optimality conditions-based single-level reductions \cite{dempe2025duality} produce non-convex MINLPs that are typically intractable at scale. To circumvent these challenges, in subsequent sections, we develop (i) a value-function-based single-level reformulation of the bilevel structure, (ii) a topology-aware sparse NN surrogate for $\eta(\bbx)$ with controlled complexity, and (iii) a mixed-integer AC physics-constrained reformulation that leverages the surrogate, enabling tractable worst-case assessment on large systems.

\subsection{Optimal Value-function Reformulation of AWLS Problem}
\label{subsec:VFR}
Under the assumption that the lower-level optimum, $\eta(\bbx)$,  is attained for every \(\bbx\in\mathcal X(k)\) (see Remark \ref{rem:feasibility}), we discuss an optimal value-function (OVF) reformulation to \textit{exactly} transform the bilevel AWLS problem \eqref{eq:awls_upper}--\eqref{eq:awls_lower} into a single-level mixed-integer nonlinear programming problem: 
\begin{subequations} 
\label{eq:value_function_reform} 
\begin{align} \max_{\bbx \in \mathcal{X}(k), \ \bby,\  \bm{\delta}^P, \bm{\delta}^Q} \quad & \sum_{i \in \mathcal{N}} \left( \delta_i^P + \delta_i^Q \right) \\ \text{s.t.} \quad & f(\bbx, \bby) \leqslant 0, \label{eq:reform_ineq} \\ 
& h(\bbx, \bby, \bm{\delta}^P, \bm{\delta}^Q) = 0, \label{eq:reform_eq} \\ 
& \sum_{i \in \mathcal{N}} \left( \delta_i^P + \delta_i^Q \right) \leqslant \eta(\bbx). \label{eq:value_constraint} 
\end{align} 
\end{subequations}

Formulation \eqref{eq:value_function_reform} maximizes the realized total load shed subject to the AC-physics feasibility constraints with line switching, while the value-function inequality enforces that the realized load shed cannot exceed the lower-level optimal $\eta(\bbx)$ associated with the chosen topology. Constraint \eqref{eq:value_constraint} is 
necessarily tight at every feasible point of 
\eqref{eq:value_function_reform}: for fixed $\bbx$, every lower-level 
feasible point satisfies 
$\sum_{i\in\mathcal N}({\delta}_i^P+{\delta}_i^Q)\geqslant \eta(\bbx)$, while 
\eqref{eq:value_constraint} enforces the reverse inequality. Hence the 
realized load shed equals $\eta(\bbx)$ and 
$(\bby,\bm{\delta}^P,\bm{\delta}^Q)$ is a lower-level minimizer. Under the 
feasibility guaranteed by Remark~\ref{rem:feasibility}, 
\eqref{eq:value_function_reform} is therefore exactly equivalent to the 
bilevel problem \eqref{eq:awls_upper}--\eqref{eq:awls_lower}.

OVF reformulations of bilevel programs are well known \cite{dempe2013bilevel}: they eliminate the nested attacker–defender structure by replacing the lower level with its value-function. This perspective lends itself to algorithmic approaches that solve the bilevel problem efficiently in a single-level framework. However, this reformulation introduces two significant challenges: (a) The AC power flow constraints in \eqref{eq:reform_ineq}--\eqref{eq:reform_eq}, when combined with discrete line-switching variables $\bbx$, create a disconnected non-convex feasible region, and (b) The constraint \eqref{eq:value_constraint} requires evaluation of $\eta(\bbx)$, which is itself the optimal value of the lower-level optimization problem \eqref{eq:awls_lower}.

We address these challenges by first applying quadratic convex relaxations with line switching to handle the nonconvex on/off AC constraints (see \cite{guo2025tightening,bestuzheva2020convex} for details), and second by developing efficient neural network (NN)–based surrogate models to approximate the value-function $\eta(\bbx)$, which enables a fully single-level reformulation as described in the following section.

\section{Neural Approximation of Value-Function } \label{sec:NN}
We approximate the value-function $\eta(\bbx)$ with a piecewise-linear (PWL) NN surrogate. As ReLU networks are PWL maps of their inputs, the trained model admits an exact MILP representation. Figure~\ref{fig:structure} depicts the overall architecture. The NN takes as inputs the topology vector $\bbx$ and the load profiles $\{\bbP^D,\bbQ^D\}$; we denote the concatenated input by $\tilde{\bbx} := [\bbx^\top, (\bbP^D)^\top, (\bbQ^D)^\top]^\top$.

\subsection{Model and training}
Consider a single-hidden-layer ReLU network
\begin{subequations}
\label{eq:NN_pf}
\begin{align}
    \bbz^{1} &= \sigma(\bbW^{1} \tilde{\bbx} + \bbb), \label{NN_pf1}\\
    \hat \eta(\bbx) &= \bbW^{2} \bbz^{1}, \label{NN_pf2}
\end{align}
\end{subequations}
where ReLU activation function $\sigma(\cdot)$ is applied elementwise and $\hat\eta(\bbx)$ denotes the NN approximation of $\eta(\bbx)$. Note, deeper NNs follow by stacking \eqref{NN_pf1} across layers; all derivations below extend verbatim.

To optimize the trainable NN parameters $\{\bbW^1, \bbW^2, \bbb \}$ and improve the accuracy of the approximation~\cite{cho2023topology}, we use the squared-error loss. Based on the training dataset, generated from solved lower-level problems \eqref{eq:awls_lower}, we can calculate the errors between the true $\eta(\bbx)$ and the approximated $\hat \eta(\bbx)$, given by
\begin{align}
    \mathcal{L}({\bbW^1, \bbW^2, \bbb})= &\| \eta(\bbx) - \hat \eta(\bbx)  \|^2_2. \label{update}
\end{align}

\subsection{Exact MILP encoding of ReLU}
Because $\sigma$ is ReLU, the map $\tilde{\bbx}\mapsto \hat\eta(\bbx)$ in \eqref{eq:NN_pf} is piecewise linear. Hence, it suffices to consider the reformulation of the ReLU functions as follows:
\begin{align}
    \bbz^1 = \sigma(\bbW^1 \tdbbx + \bbb) =  (\bbW^1 \tdbbx + \bbb) \odot \bbbeta \label{eq:prod}
\end{align}
where $\bbbeta$ is an auxiliary binary vector indicating the ReLU activation status, while $\odot$ denotes the entry-wise vector multiplication.
This representation allows us to adopt the well-known \textit{big-M tightening method} for the following equivalent linear reformulation~\cite{griva2009linear}:

For the $k$-th ReLU activation $z^{1}_k$ in \eqref{NN_pf1}, 
let $\underline M_k \leqslant 0$ and $\bar M_k \geqslant 0$ denote valid lower and upper 
bounds on the pre-activation value $\hat z^1_k$, i.e., 
$\underline M_k \leqslant \hat z^1_k \leqslant \bar M_k$. After determining these 
bounds and denoting the input in \eqref{NN_pf1} by 
$\hat {\bbz}^{1} = \bbW^1 \tdbbx + \bbb$, the big-M method 
gives the following exact MILP encoding of the ReLU relation 
$z^1_k=\max\{0,\hat z^1_k\}$ and hence of the value-function approximation 
\eqref{eq:NN_pf}:
\begin{subequations}\label{NN_VF}
\begin{align}
    \hat \bbz^{1} &= \bbW^{1} \tdbbx + \bbb,\\
    \hat {z}^{1}_k &\leqslant z^{1}_k \leqslant \hat {z}^{1}_k - \underline M_k (1- \beta_k),\label{beta1}\\
    0 &\leqslant  z^{1}_k \leqslant \bar M_k  \beta_k,\label{beta2}\\
    \hat \eta(\bbx) &= \bbW^2 \bbz^{1}.
\end{align}
\end{subequations}
Here, $\beta_k\in\{0,1\}$ indicates the ReLU activation status. If $\hat z^1_k>0$, then $\beta_k=0$ is infeasible (it would force $z^1_k=0$ while the lower bound requires $z^1_k\geqslant \hat z^1_k>0$), so $\beta_k=1$ and $z^1_k=\hat z^1_k$. If $\hat z^1_k<0$, then $\beta_k=1$ is infeasible (it would force $z^1_k\leqslant \hat z^1_k<0$ while $z^1_k\geqslant 0$), so $\beta_k=0$ and $z^1_k=0$. When $\hat z^1_k=0$, both values of $\beta_k$ are feasible and yield $z^1_k=0$. Thus, the formulation enforces $z^1_k=\max\{0,\hat z^1_k\}$ exactly for valid pre-activation bounds $\underline M_k,\bar M_k$; tighter bounds yield a stronger continuous relaxation~\cite{anderson2020strong}.

\begin{figure}[t!]
	\begin{center}
		\includegraphics[scale=0.33]{Figures/NN_v4.png}
		\caption{\small The proposed NN model predicts the approximate optimal load shed $\hat \eta(\bbx)$ for any given topology $\bbx$ and active/reactive loads.} 
        \label{fig:structure}
	\end{center}
\end{figure}

\subsection{Preliminary NN Surrogate for the AWLS}\label{sec:prelim}
The NN surrogate $\hat{\eta}(\bbx)$ in \eqref{NN_VF} provides an approximation of the true $\eta(\bbx)$ for the upper-level problem \eqref{eq:awls_upper}. Substituting $\hat{\eta}(\bbx)$ for $\eta(\bbx)$ yields the single-level problem:
\begin{align}
\label{eq:nn_awls}
    \max_{\bbx \in \mathcal{X}(k)} \quad & \hat \eta(\bbx).
\end{align}
This formulation admits an exact MILP representation because the ReLU activation functions can be encoded via the big-$M$ constraints in \eqref{NN_VF}.
We emphasize that \eqref{eq:nn_awls} is a purely surrogate-based upper-level search: during this optimization, the AC power-flow feasibility variables and constraints are not included explicitly, because their effect is represented only through the learned value-function approximation $\hat{\eta}(\bbx)$. This makes \eqref{eq:nn_awls} computationally attractive and provides a tractable data-driven baseline in the spirit of~\cite{aquino2023comparing}. However, if $\hat{\eta}(\bbx)$ misranks topologies, especially under distribution shift, direct maximization of \eqref{eq:nn_awls} may select a topology whose predicted worst-case load shed is inaccurate when re-evaluated by the original lower-level AC operator-response model. In the next section, we address this limitation by coupling the value-function surrogate with physics-based feasibility information through the OVF reformulation.

\section{Physics-constrained Sparse NN Surrogate} \label{sec:approx}
While the surrogate maximization in \eqref{eq:nn_awls} is computationally attractive, it is purely data-driven and does not enforce the power-flow feasibility constraints. Consequently, it can select topologies whose predicted load shed approximates the true value-function $\eta(\bbx)$, leading to physically inaccurate rankings. To address this limitation, we adopt a physics-constrained OVF reformulation (see Section~\ref{subsec:VFR}) that anchors the surrogate to grid physics by coupling the objective with the feasibility set and a value-function inequality.

Specifically, we introduce a physics-constrained neural surrogate (PCNN) embedded in the OVF reformulation in \eqref{eq:value_constraint}: any candidate $\bbx$ must admit $(\bby,\mathbf{ls}^P,\mathbf{ls}^Q)$ satisfying the power-flow constraints, and the realized load shed is tethered to the surrogate through an explicit value-function constraint. To retain scalability, we further develop a sparse, area-partitioned NN architecture that reduces the surrogate size—and thus the number of MILP binaries—approximately linearly with system size. We will present these two contributions in order.

\subsection{Physics-constrained NN (PCNN) Surrogate}\label{sec:PCNN}
The purely data-driven surrogate maximization in \eqref{eq:nn_awls} is a statistical prediction and can misrank topologies by
(i) \emph{overestimation}, when $\hat\eta(\bbx)$ is unrealistically high for a topology $\bbx$ (the surrogate selects an unachievable ``worst case''), and
(ii) \emph{underestimation}, when $\hat\eta(\bbx)$ is too low (the surrogate fails to reveal the true worst case). To tackle both failure modes, we anchor the surrogate to grid physics via the OVF reformulation \eqref{eq:value_function_reform}, yielding the following physics-constrained neural surrogate (PCNN): 
\begin{subequations} 
\label{eq:pcnn_awls} 
\begin{align} \max_{\bbx \in \mathcal{X}(k), \ \bby,\  \bm{\delta}^P, \bm{\delta}^Q, s^{\delta}} \quad & \sum_{i \in \mathcal{N}} \left( \delta_i^P + \delta_i^Q \right) - \lambda\cdot s^{\delta} \\ \text{s.t.} \quad & f(\bbx, \bby) \leqslant 0, \label{eq:reform_ineq_2} \\ 
& h(\bbx, \bby, \bm{\delta}^P, \bm{\delta}^Q) = 0, \label{eq:reform_eq_2} \\ 
& \sum_{i \in \mathcal{N}} \left( \delta_i^P + \delta_i^Q \right) \leqslant \hat \eta(\bbx) + s^{\delta}, \label{eq:value_constraint_2} \\
& 0 \leqslant s^{\delta} \leqslant \overline{s},
\end{align} 
\end{subequations}
where $s^{\delta}$ is the non-negative slack variable in the value–function constraint. The physics constraints \eqref{eq:reform_ineq_2}–\eqref{eq:reform_eq_2} enforce AC feasibility with line switching, so any feasible $(\bby,\bm{\delta}^P,\bm{\delta}^Q)$ corresponds to a physically achievable total shed $\sum_i(\delta_i^P+\delta_i^Q)$; this prevents the optimizer from realizing NN overestimates of infeasible load shed. The slack $s^{\delta}$ mitigates surrogate \emph{underestimation}: if the physics-feasible shed exceeds $\hat\eta(\bbx)$, $s^{\delta}$ absorbs the gap (penalized by $\lambda$) to avoid truncating the true worst case.

We note that the linear NN output is used primarily as a surrogate ranking score over candidate topologies, rather than as an independently certified physical load-shedding value. Thus, possible conservative predictions, including hypothetical negative values, are interpreted as surrogate underestimation rather than as physical load-shedding quantities. In the PCNN formulation, such underestimation is handled through the slack variable $s^\delta$ in \eqref{eq:value_constraint_2}. Within the prescribed slack cap, $s^\delta$ allows the physics-feasible load shedding to exceed $\hat{\eta}(\bbx)$, while the realized load shedding remains nonnegative and physically constrained through $(\bm{\delta}^P,\bm{\delta}^Q)$ and \eqref{eq:reform_ineq_2}-\eqref{eq:reform_eq_2}.
Since Remark~\ref{rem:feasibility} provides an upper bound on the true value-function, a sufficiently large non-negative $M^s$ accounting for negative NN predictions yields a valid slack bound: $s^{\delta} \leqslant \overline{s} := \sum_{i\in\mathcal N}\bigl(P_i^D+Q_i^D\bigr) + M^s.$
When available, tighter calibrated caps on $s^{\delta}$ further strengthen the PCNN formulation.

\begin{remark} \hspace{6cm}\\
        \textbf{(a)} \underline{Exact surrogate:} If $\hat{\eta}(\bbx)\equiv \eta(\bbx)$ for all $\bbx\in\mathcal X(k)$, then the global optimum of both PCNN in \eqref{eq:pcnn_awls} and direct surrogate maximization in \eqref{eq:nn_awls} coincides with the bilevel AWLS problem in \ref{subsec:awls_prob} in both optimal value and maximizers. For PCNN, this occurs with $s^{\delta\star}=0$ at optimality. \\
        \textbf{(b)} \underline{Overestimating surrogate:} In PCNN, if the surrogate is a pointwise upper bound, that is, $\hat{\eta}(\bbx)\geqslant \eta(\bbx)$ for all $\bbx\in\mathcal X(k)$ and $\lambda > 0$, then there exists an optimal PCNN solution with $s^{\delta\star}=0$. However, topology misranking may still yield suboptimal solutions to the bilevel AWLS problem despite zero slack. 
        \textbf{(c) }\underline{Rank-preserving surrogate (under- or overestimation):} If $\hat{\eta}(\bbx)$ preserves the ranking of $\eta(\bbx)$ on $\mathcal X(k)$ --- i.e., $\eta(\bbx_a) \geqslant \eta(\bbx_b) \iff \hat{\eta}(\bbx_a)\geqslant \hat{\eta}(\bbx_b)$ for all $\bbx_a,\bbx_b \in \mathcal{X}(k)$ --- then direct surrogate maximization in \eqref{eq:nn_awls} recovers the global maximizer $\bbx^*$ of the bilevel AWLS problem. However, PCNN requires careful $\lambda$ selection to recover the global optimum, as the penalty term $-\lambda \cdot s^{\delta}$ can alter topology rankings.
\end{remark}

\begin{figure}[t!]
	\begin{center}
		\includegraphics[width=0.9\linewidth]{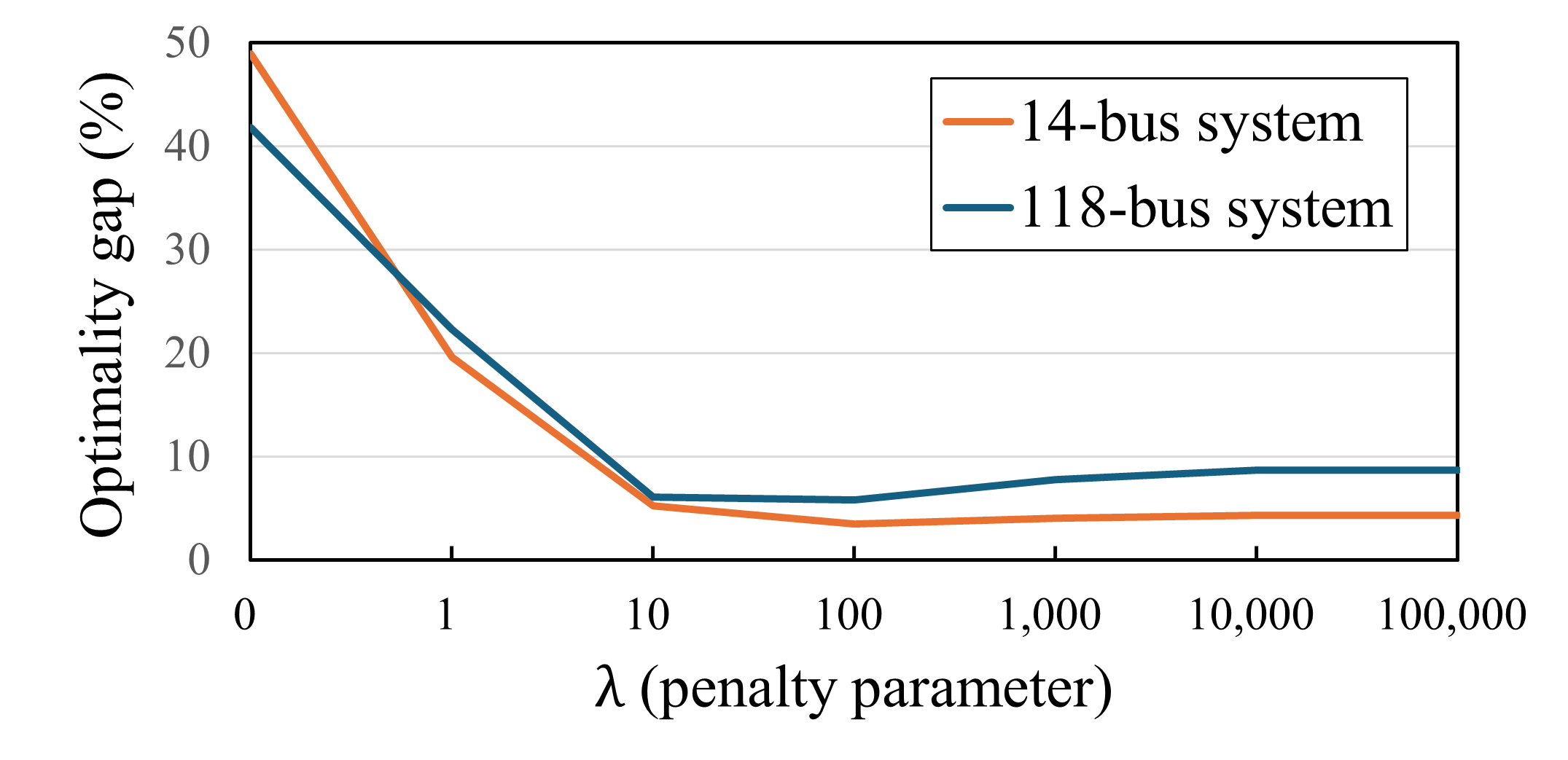}
		\caption{Optimality gap as a function of the slack-regularization parameter $\lambda$ on the IEEE 14- and 118-bus systems.} \label{SelectionLambda}
	\end{center}
\end{figure}

\begin{remark}[Selecting the penalty $\lambda$]\label{rmk:slack}
The parameter $\lambda > 0$ trades off fidelity to the surrogate constraint and flexibility to correct underestimation. The term $-\lambda \cdot s^{\delta}$ suppresses slack: very large $\lambda$ effectively enforces $s^{\delta} \rightarrow 0$ (risking truncated worst cases), whereas very small $\lambda$ permits excessive slack (degrading optimality). We tune $\lambda$ by validation; on the IEEE tests the optimality gap (defined in Section~\ref{subsec:benchmark}) exhibits insensitivity plateaus for $\lambda\!\in\![10,100]$ (see Fig.~\ref{SelectionLambda}), and we fix $\lambda$ in this range. If the objective is non-normalized, a stable choice can be a scaled parameterization $\lambda=\bar\lambda/\overline s$ with $\bar\lambda\in[10,100]$.
\end{remark}

The PCNN formulation in \eqref{eq:pcnn_awls} is a nonconvex mixed-integer nonlinear program (MINLP) due to the AC power-flow constraints \eqref{eq:reform_ineq_2}–\eqref{eq:reform_eq_2} coupled with the binary line-switching variables $\bbx$ and the mixed-integer linear block due to NN embedding \eqref{NN_VF}. Solving this MINLP to global optimality is generally intractable; we apply on/off quadratic convex relaxations of the AC physics (see \cite{bestuzheva2020convex, guo2025tightening} for details), yielding a mixed-integer convex program. Even though this relaxed PCNN is more computationally demanding than the purely NN-based surrogate maximization in \eqref{eq:nn_awls}, it typically returns higher-quality and physically reliable topologies, as demonstrated in our numerical studies.

\subsection{Sparsifying NN Surrogates using Area Partitioning} \label{sec:clustering}
While the PCNN in \eqref{eq:pcnn_awls} enforces power‐flow feasibility, its quality and tractability hinge on the accuracy of the value-function surrogate $\hat{\eta}(\bbx)$. Approximating $\eta(\bbx)$ at scale is challenging due to model complexity and training data required to capture highly-variable system-wide responses. To address this, we propose a sparse, topology-aware (area-partitioned) NN architecture. Based on the core intuition that load-shedding decisions are dominated by local grid characteristics, we partition the system to create smaller, specialized NNs for each area. This approach significantly reduces the number of ReLU neurons, and hence MILP binaries approximately linearly with system size, while preserving fidelity to localized load-shedding behavior—producing a surrogate with a more tractable MILP formulation.

\begin{figure}[t!]
	\begin{center}
		\includegraphics[scale=0.25]{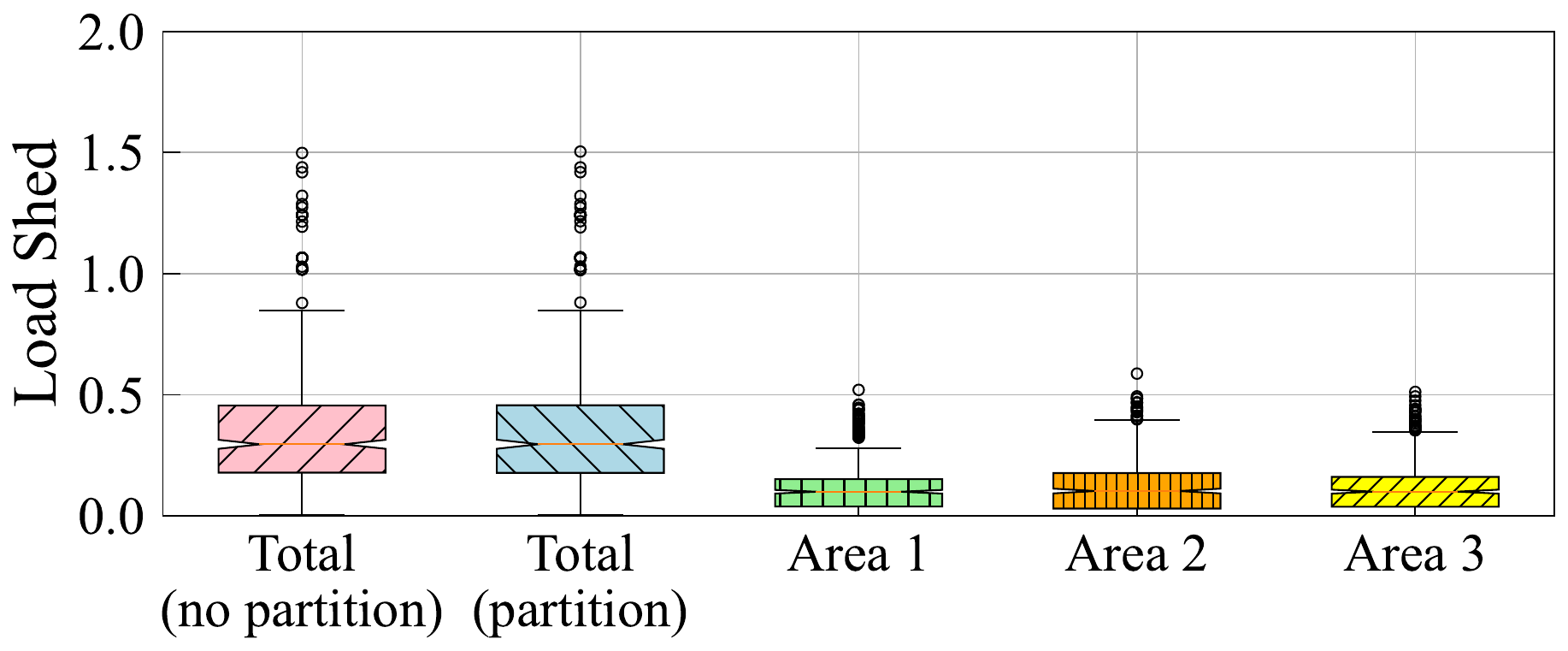}
        \caption{\small Optimal lower-level (NLP) solutions for IEEE 118 system under randomly sampled line interdictions. Distributions of total load shed are shown for (i) full system without partitioning (red), (ii) simple sum of the three area sheds (blue), and (iii) each area individually. Per-area distributions are significantly narrower, indicating spatial localization, while the area-sum (blue) closely tracks the full-system shed (red), confirming that area partitioning preserves aggregate shed.}
        \label{fig:motivation}
	\end{center}
\end{figure}

To motivate partitioning, we examine the variability of the target value-function $\eta(\bbx)$. Figure~\ref{fig:motivation} reports distributions of $\eta(\bbx)$ computed from locally optimal solutions of the lower-level NLP \eqref{eq:awls_lower} for 546 randomly sampled ($k=5$) interdictions on the IEEE 118-bus system at nominal load. The full-system value (red box) exhibits large dispersion across topologies, whereas the three per-area sheds are much tighter, indicating strong spatial localization. Crucially, the simple sum of the individually solved per-area sheds (blue box) closely matches the full-system shed (red box), confirming that aggregation preserves the total $\eta(\bbx)$. This variance reduction simplifies the learning task and enables compact, area-wise NN surrogates.

\begin{figure}[t!]
	\begin{center}
		\includegraphics[width=0.97\linewidth]{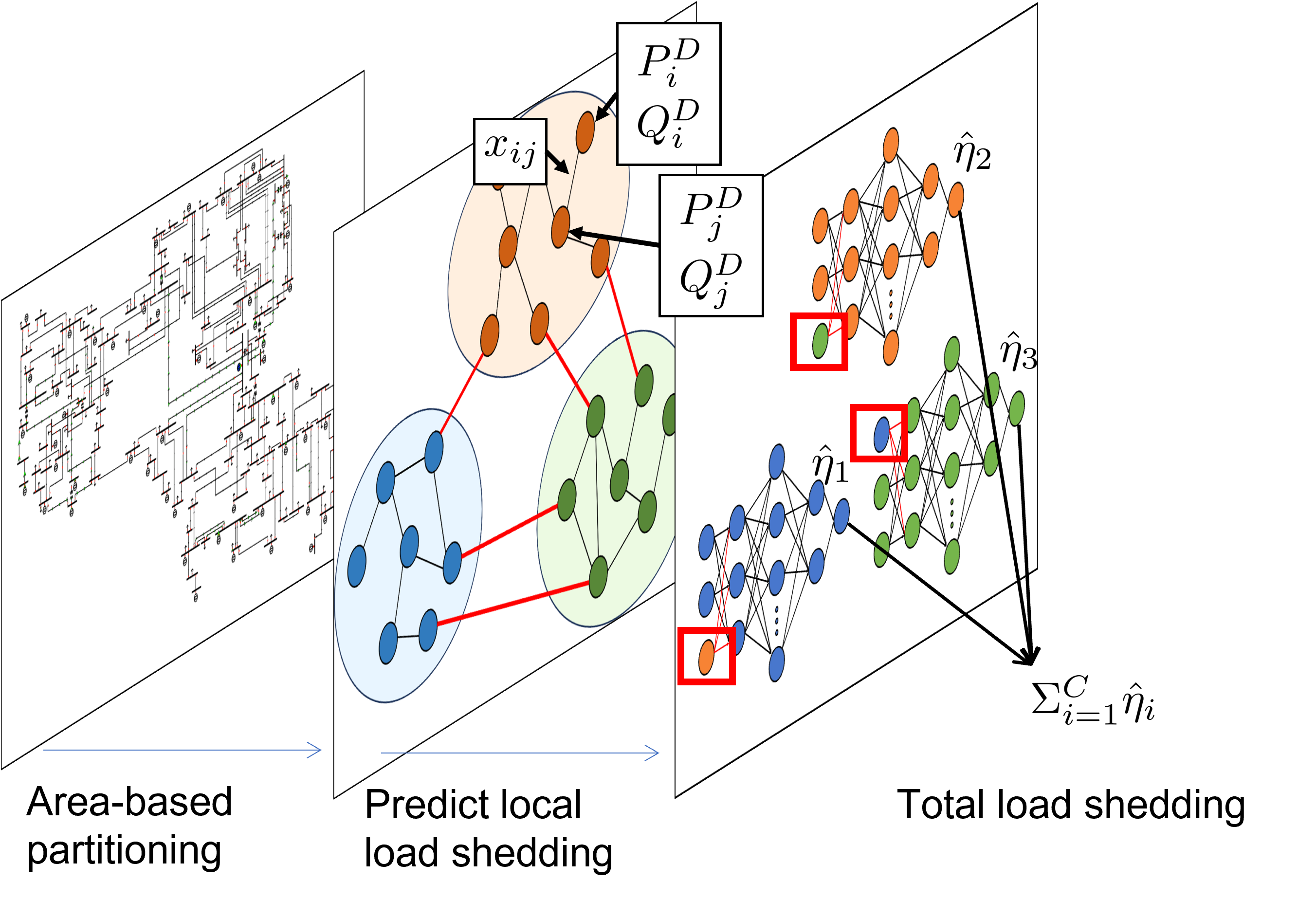}
        \caption{\small Our architecture partitions the system into electrically coherent areas via spectral clustering (three shown for illustration), then uses per-area NNs with tie-line inputs to predict local load shedding, summing the results for the total system prediction.}
        \label{structure_part}
	\end{center}
\end{figure}

We implement this partitioning strategy through the sparse NN architecture shown in Fig.~\ref{structure_part}. The full power grid without any contingency (left) is first divided into $C$ distinct areas. We then design a compact per-area NN for each area $i$ to predict its local load-shed contribution $\hat{\eta}_i(\bbx)$ for a given $\bbx$. To optimize this modular structure, the number of hidden neurons for area $i$ is budgeted according to the variability of its historical (or simulated) local load shed, assigning more neurons to higher-variance areas. Each per-area NN primarily uses local features (e.g., loads and intra-area line statuses) and augments them with a small set of shared tie‑line features (end-bus states and line impedance for the boundary lines) to capture inter-area dependencies. All area-specific NNs are trained jointly so that $\sum_{i=1}^{C}\hat{\eta}_i(\bbx)$ accurately approximates the system-wide load shed $\eta(\bbx)$. 

\begin{remark}
    By construction, the NN's weight matrices $W^{(\ell)}$ are nearly block 
diagonal when inputs and hidden neurons are ordered by area, with only 
sparse off-diagonal entries for tie-line features. Let area $i$ use 
$d_i=\Theta(N')$ local input features and $h_i=\kappa_i d_i$ hidden 
neurons with $\kappa_i\leq \kappa_{\max}$ independent of $N$. With 
$C=N/N'$ areas,
\[
\sum_{i=1}^{C} d_i h_i
\leq \kappa_{\max}\sum_{i=1}^{C}(N')^2
= \kappa_{\max}\cdot \frac{N}{N'}\cdot (N')^2
= O(N\cdot N').
\]
We enforce size-constrained area partitioning so that $N'=\Theta(1)$, 
i.e., $\max_i d_i\leq d_{\max}$, which gives 
$\sum_{i=1}^{C}d_i h_i=O(N)$. Each area-specific NN also uses only a 
bounded number of tie-line features from incident boundary branches, 
with each retained boundary branch contributing a fixed number of 
line-status and electrical-parameter features. Hence, the off-diagonal 
boundary couplings add $\sum_{i=1}^{C}O(1)=O(C)=O(N)$ parameters. With 
fixed depth, bias terms and hidden-to-output connections contribute 
$O(N)$ parameters, resulting in overall linear scaling.
\end{remark}

\noindent
\textbf{Spectral Partitioning} An appropriate area partition is critical to ensure our decomposition is physically meaningful. We employ \textit{spectral clustering} using electrical distance as the graph-edge weight, based on the rationale that electrically distant areas are more self-sufficient, with high-impedance tie-lines having minimal impact on local load shed~\cite{sanchez2014hierarchical}. In particular, each branch $(i,j) \in \mathcal{L}$ is assigned the weight $\lambda_{ij}:=1/|z_{ij}|$, where $z_{ij}$ is the series impedance magnitude, so that buses joined by a low-impedance line are “electrically close.” Construct the corresponding weighted Laplacian matrix $\bbLambda\in\mathbb{R}^{N\times N}$, 
\begin{align}\nonumber
    \bbLambda_{ij} &= \begin{cases}
    \textstyle \sum_{j=1}^n \lambda_{ij}, & \text{if $ i= j$,}\\
    -\lambda_{ij}, & \text{if $(i,j)\in \mathcal{L}$,}\\
    0, & \text{otherwise.}
    \end{cases} 
\end{align}
Let $\widehat{d}$ denote the spectral embedding dimension, and let 
$u_1,\ldots,u_{\widehat{d}}$ be the eigenvectors associated with the 
$\widehat{d}$ smallest non-trivial eigenvalues of $\bbLambda$. Each bus 
$i$ is embedded into the spectral coordinate vector 
$r_i:=(u_1(i),\ldots,u_{\widehat{d}}(i))$, formed from the $i$-th 
entries of these eigenvectors. Applying the $k$-means algorithm to 
$\{r_i\}_{i\in\mathcal N}$ partitions the buses into $C$ electrically 
coherent areas, where $C$ need not equal $\widehat{d}$. This yields clusters with strong intra-area electrical coupling and weak inter-area ties, yielding the sparse connectivity needed for our block-sparse NN architecture with minimal boundary corrections. See~\cite{cho2024data} for more discussion. 

Leveraging spectral partitioning and block-sparse connectivity, the architecture attains linear parameter growth while preserving accuracy and data efficiency, yielding a scalable surrogate $\hat{\eta}(\bbx)$. We validate these claims empirically in the next section.

\section{Numerical Studies} 
\label{sec:results}
We validated the proposed sparse NN–based approach for the AWLS problem on the IEEE 14- and 118-bus test systems~\cite{IEEE_case_ref}. NN Training was performed in PyTorch (Adam optimizer) on a commodity workstation (Intel\textsuperscript{\textregistered} CPU @ 2.70\,GHz, 32\,GB RAM, NVIDIA\textsuperscript{\textregistered} RTX 3070 Ti, 8\,GB VRAM). All optimization models were implemented in JuMP~\cite{lubin2023jump}; Gurobi 11.0.2 solved the mixed-integer linear and convex programs, and IPOPT 3.14.4 solved the continuous nonlinear programs~\cite{gurobi,wachter2006implementation}.

We generate training datasets comprising 1,000 active and reactive load profiles (each OPF-feasible for the original topology) for the 14- and 118-bus systems, with 50 and 428 distinct topologies, respectively. For each system, we define critical line damage sets (8 and 15 lines) vulnerable to $k=3$ simultaneous outages, excluding topologies that cause islanding.
For each system, we enumerate all $k=3$ outage combinations under the nominal load condition, solve the lower-level load-shedding problem for each topology, and rank the resulting topologies by their load-shedding values. We then take the union of the lines appearing in the top 100 highest-impact topologies as the vulnerable candidate set, yielding 8 candidate lines for the 14-bus system and 15 candidate lines for the 118-bus system. This procedure reflects a practical setting in which operators first identify vulnerable corridors offline and then use the proposed surrogate-based AWLS solver for fast evaluation under varying load profiles.
Load profiles are generated by sampling uniformly from high, medium, and low loading ranges to ensure representative coverage of operating conditions.

We solve the lower-level NLP problem (Eq. \eqref{eq:awls_lower}) for each topology-load profile sample using IPOPT to obtain locally-optimal load shed values as training labels. The NN is trained via backpropagation with loss function \eqref{update} for 1,000 epochs using a learning rate of $2.5\times10^{-3}$. We employ a 90\%-10\% train-test split, with all reported performance metrics computed exclusively on the test set to ensure unbiased evaluation.

\begin{figure}[t!]
	\centering
	\subfloat[14-bus system]{\includegraphics[scale=0.27]{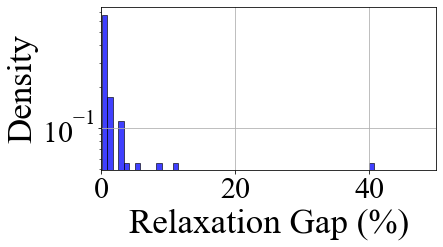}}
	\quad
	\subfloat[118-bus system]{\includegraphics[scale=0.27]{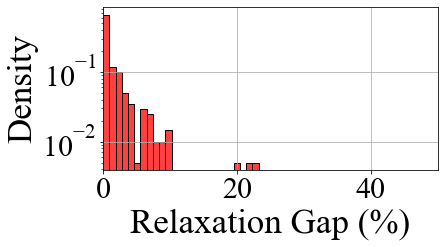}}
	\caption[]{\small Distribution of the relaxation gaps for (a) 14-bus and (b) 118-bus systems under the nominal load condition.}
    \label{fig:QC_relax}
\end{figure}

We set the big-M bounds $(\bar{M}_k, \underline{M}_k)$ to (-100, 100) for the MILP reformulation of ReLU units, as empirical analysis confirms the pre-activation values $\hat{\bbz}^1$ remain within this range. For the AC power flow constraints in the PCNN surrogate \eqref{eq:pcnn_awls}, we implement an on/off version of the convex relaxation model via PowerModels~\cite{8442948}.
This convex formulation reduces solution times by approximately 90\% compared to the nonconvex AC model, while remaining tight in most tested instances.
Figure~\ref{fig:QC_relax} shows the distribution of the relaxation gap for a single nominal load profile, computed across all considered outage topologies. Specifically, for each topology, we compare the lower-level objective value obtained from the relaxation against that obtained from the original nonconvex AC lower-level model. The relaxation is tight for most topologies, with 98.0\% and 99.3\% of the considered topologies for the 14- and 118-bus systems, respectively, exhibiting relaxation gaps below 10\%. This indicates that the convex relaxation model typically preserves the relevant topology ranking well, although a few outliers exhibit larger gaps.


\subsection{On Benchmarking the AWLS Problem}
\label{subsec:benchmark}

\noindent
\textbf{Enumerated AWLS Benchmark} 
In the absence of tractable methods that are globally certified for solving bilevel AWLS with nonconvex AC constraints (Section~II-A), we adopt an enumeration-based AC-NLP benchmark. Specifically, under the budget $k$, we enumerate $\mathbf{x}\in\mathcal{X}(k)$ over the considered connected candidate-topology set and solve each lower-level problem $\eta(\mathbf{x})$ to \emph{local} optimality. This benchmark fully achieves the upper-level combinatorial search over the considered topology set, while the remaining lack of global certification comes from the nonconvexity of the lower-level AC load-shedding problem.
Although computationally expensive, the resulting $\mathbf{x}^\star\in\arg\max_{\mathbf{x}\in\mathcal{X}(k)}\eta(\mathbf{x})$ with value $\eta(\mathbf{x}^\star)$ provides a rigorous benchmark.
For the 118-bus system with 15 candidate lines and $k=5$ simultaneous outages, full enumeration requires solving $\binom{15}{5}=3003$ lower-level nonlinear load-shedding problems for each load-profile instance. With an average IPOPT solution time of approximately 24 seconds per topology, this corresponds to roughly 20 hours for a single load profile.
We report the optimality gap $\frac{|\hat{\eta}-\eta(\mathbf{x}^\star)|}{\eta(\mathbf{x}^\star)}\cdot 100$, where $\hat{\eta}$ is the optimal objective of the single-level surrogate-embedded formulation, as explained below.

For each load profile in the test set, we solve the surrogate-embedded single-level problem in \eqref{eq:nn_awls} and \eqref{eq:pcnn_awls}, using the pool-based refinement strategy (in Remark \ref{rmk:pool}), to obtain a topology $\hat{\bbx}$. We then compute the realized load shed $\hat{\eta} \equiv \eta(\hat{\mathbf{x}})$, enabling a fair assessment against the enumerated benchmark value $\eta(\mathbf{x}^\star)$.

\begin{remark}{(Pool-based refinement)}
\label{rmk:pool}
To hedge against NN surrogate misranking, we apply a pool-based refinement across the two \emph{single-level} surrogates --- the direct NN maximization \eqref{eq:nn_awls} and the PCNN formulation \eqref{eq:pcnn_awls}. During each solve, we retain up to $K=10$ top-scoring topology incumbents $\{\bbx^{(1)},\ldots,\bbx^{(K)}\}$ returned by the MIP solver. For each candidate $\bbx^{(m)}$, we verify it by re-solving the lower-level NLP \eqref{eq:awls_lower} with $\bbx=\bbx^{(m)} \ \forall m$ to local optimality and then select the final topology as $\hat{\bbx}=\arg\max_m \eta(\bbx^{(m)})$. This preserves the computational gains of the surrogates while returning improved worst-case topology solutions.
\end{remark}

\subsection{Value-function Approximation Quality}

We first validate the NN surrogate by comparing predicted value-functions $\hat{\eta}(\mathbf{x})$ against optimal values $\eta(\mathbf{x})$ obtained from locally optimal solutions of the continuous lower-level problem~\eqref{eq:awls_lower} over randomly sampled topologies. For each topology–load profile pair, the normalized prediction error is
$
\frac{|\hat{\eta}(\mathbf{x})-\eta(\mathbf{x})|}{\eta(\mathbf{x})}\cdot 100.
$ \\

\begin{figure}[t!]
	\centering
	\subfloat[14-bus system]{\includegraphics[scale=0.27]{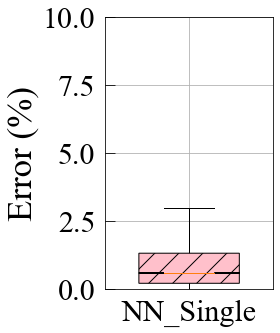}}
	\quad
	\subfloat[118-bus system]{\includegraphics[scale=0.27]{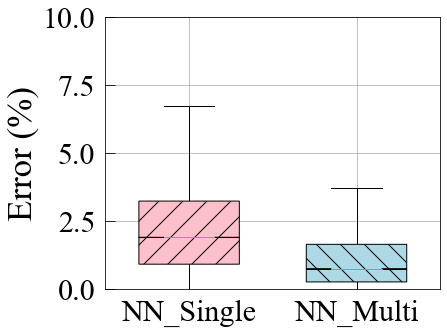}}
	\caption[]{\small Comparisons of the prediction errors (\%) in approximating the load sheds for NN_Single and NN_Multi models.}\label{Approx_performance}
\end{figure}

\renewcommand{\arraystretch}{1.2}
\begin{table}[t!]
\caption{\small Comparison of rank correlation coefficients (Kendall's $\tau$, Spearman's $\rho$) between the true and predicted value-functions; each entry lists the pair $\tau, \rho$.}
\begin{center}
\begin{tabular}{c|c|c|c}
\Xhline{3\arrayrulewidth}
& \textbf{14-bus} & \multicolumn{2}{c}{\textbf{118-bus}} \\
& NN\_Single & NN\_Single & NN\_Multi \\ \hline
min & 0.786, 0.885 & 0.445, 0.616 & 0.584, 0.776 \\
avg & 0.920, 0.986 & 0.871, 0.970 & 0.912, 0.984 \\
max & 0.965, 0.997 & 0.953, 0.996 & 0.978, 0.999 \\
\Xhline{3\arrayrulewidth}
\end{tabular}\label{rank_performance}
\end{center}
\end{table}

\renewcommand{\arraystretch}{1.2}
\begin{table*}[t!]
\caption{\small AWLS problem comparisons of the NN and PCNN surrogates on the 14- and 118-bus systems for NN\_Single and NN\_Multi (bolded numbers marking the better-performing surrogate).
}
\begin{center}
\begin{tabular}{c|c|c|c|c|c|c|c}
\Xhline{3\arrayrulewidth}
  &  & \multicolumn{2}{c|}{14-bus} & \multicolumn{4}{c}{118-bus} \\ \cline{3-8}
  & & \multicolumn{2}{c|}{NN\_Single} & \multicolumn{2}{c|}{NN\_Single} & \multicolumn{2}{c}{NN\_Multi}\\ \cline{3-8}
  & & NN & PCNN & NN & PCNN & NN & PCNN \\ \hline
  \multirow{3}{*}{{\makecell[c]{Optimality\\ Gap (\%)}}}  & min &  0\%    &  0\%     & 0\%    &   0\% & 0\% & 0\%\\
  & avg &  4.3\%    &  \textbf{3.5\%}     &  8.7\%   &   \textbf{5.8\%} & 7.6\%   &   \textbf{3.7\%}\\
  & max &  48.9\%    &   19.6\%    & 41.8\%    &   22.3\% & 34.5\%    &   18.6\%\\ \hline
\multirow{3}{*}{{\makecell[c]{Computation\\Time (s)}}} & min  & 0.009 s  &  0.181 s   & 0.67 s   &  13.2 s & 0.65 s   &  9.6 s\\
  & avg &  \textbf{0.01 s}   &  0.214 s    & \textbf{0.93 s}   &  16.4 s & \textbf{0.89 s}   &  12.8 s\\
  & max &  0.012 s   &  0.283 s    & 1.12 s   &  19.3 s & 1.08 s   &  14.3 s\\
\Xhline{3\arrayrulewidth}
\end{tabular}\label{WLS_Follower_Leader}
\end{center}
\end{table*}

\noindent
\textbf{14-bus accuracy}
On the 14-bus system, we train a single neural network (NN\_Single) with two hidden layers of 50 neurons each. Figure~\ref{Approx_performance}(a) shows the distribution of the normalized errors, where the midline, box, and horizontal bars represent the median, quartiles (Q1–Q3), and the range within 1.5×IQR, respectively. NN\_Single achieves a median error of $0.58\%$ and a maximum error of $3.0\%$ across randomly sampled topologies and load profiles, confirming accurate value-function approximation for the smaller system. \\ 

\noindent
\textbf{118-bus scalability: NN\_Single vs.\ NN\_Multi}
For the 118-bus system, we compare NN\_Single against an area-based modular approach (NN\_Multi) from Section~\ref{sec:clustering}. We partition the grid into five electrically coherent areas using spectral clustering with $k$-means on the top three Laplacian eigenvectors. NN\_Multi allocates neurons to areas in proportion to local load-shed variability and augments per-area inputs with tie-line features to capture boundary effects. Both models use two hidden layers, each with 100 neurons.

Figure~\ref{Approx_performance}(b) indicates that NN\_Single’s accuracy degrades with system size (median error $\approx 2.0\%$, maximum error $\approx 7.0\%$), while NN\_Multi maintains median errors below $0.75\%$ and maximum errors around $3.5\%$, with dispersion comparable to the 14-bus case. This demonstrates that exploiting spatial localization via a block-sparse, partitioned architecture preserves accuracy while scaling to larger systems. \\

\noindent
\textbf{Topology ranking performance}
For accurate solutions for the AWLS problem, it is important that the NN surrogate predicts the correct ranking of interdicted topologies, where a higher rank corresponds to greater load shed. To quantify the performance in rank ordering, we report topology rank correlation metrics, Kendall’s $\tau$ and Spearman’s $\rho$, in Table~\ref{rank_performance}, between the true value-function $\eta(\bbx)$ and its NN-based approximations. Here, values closer to $1$ indicate nearly exact topology ranking (and $-1$ indicates reversed ranking), so higher is better for both measures. Both NN surrogates achieve high rank correlation with the ground truth, but the area-partitioned NN\_Multi consistently outperforms NN\_Single on the 118-bus system. Specifically, NN\_Multi improves the average Kendall’s $\tau$ from $0.871$ to $0.912$ and the minimum $\tau$ from $0.445$ to $0.584$, which corresponds to about 10\% increase in correctly ordered topology pairs, along with similar gains for Spearman’s $\rho$. This shows that NN\_Multi improves the robustness of topology ranking under challenging load scenarios.

In summary, by exploiting local information through a sparse, area-partitioned design, NN\_Multi achieves superior accuracy on both 14-bus and 118-bus systems, while ensuring scalability for NN-surrogate-embedded optimization formulations, as demonstrated in the following sections.

\subsection{NN-based AWLS Approximation}\label{subsec:NN_AWLS}
We compare single-level surrogate formulations, direct NN maximization \eqref{eq:nn_awls} and the proposed PCNN \eqref{eq:pcnn_awls}, against the enumerated AWLS benchmark described in Section~\ref{subsec:benchmark}. \\

\noindent
\textbf{Performance of PCNN surrogate} \\ 
Table~\ref{WLS_Follower_Leader} summarizes the key performance metrics, revealing a clear accuracy-efficiency trade-off. 
Table~\ref{WLS_Follower_Leader} evaluates the final AWLS solution quality over test load profiles after solving the NN- or PCNN-embedded optimization. Thus, the optimality gaps can be larger than the prediction errors in Fig.~\ref{Approx_performance} when surrogate errors lead to misranking of high-impact topologies.
Using the NN\_Single value-function approximation, the PCNN surrogate consistently achieves superior optimality gaps compared to the direct NN surrogate: 3.5\% versus 4.3\% for the 14-bus system, and 5.8\% versus 8.7\% for the 118-bus system. This improved accuracy comes at a computational cost, as PCNN requires longer solution time (0.01 s vs. 0.21 s for 14-bus), though both methods remain practical with sub-minute solve times.

\begin{figure}[t!]
	\centering
	\subfloat[NN surrogate (14-bus)]{\includegraphics[scale=0.27]{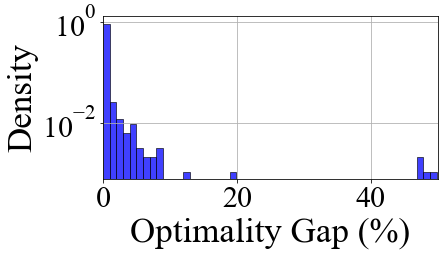}}
	\quad
	\subfloat[PCNN surrogate (14-bus)]{\includegraphics[scale=0.27]{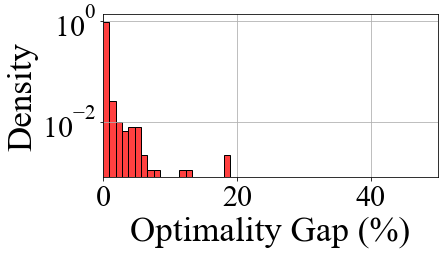}}
    \quad
    \subfloat[NN surrogate (118-bus)]{\includegraphics[scale=0.27]{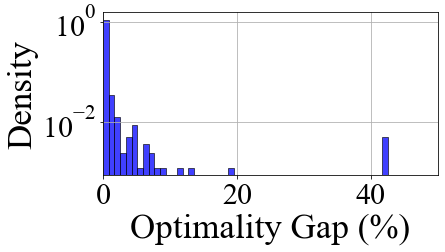}}
	\quad
	\subfloat[PCNN surrogate (118-bus)]{\includegraphics[scale=0.27]{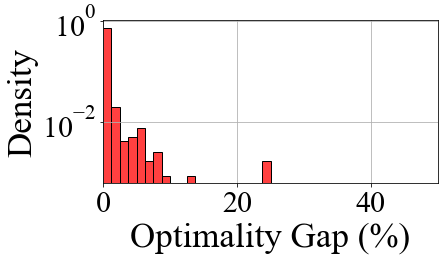}}
	\caption[]{\small Optimality gap comparisons for surrogate-embedded AWLS using NN\_Single as the value-function: (a–b) 14-bus and (c–d) 118-bus systems; panels (a,c) show the NN surrogate, and panels (b,d) show the PCNN surrogate. }\label{fig:WLS_14}
\end{figure}

Distributional analysis of results further demonstrates PCNN surrogate's improved reliability. Figure~\ref{fig:WLS_14} shows the distribution of optimality gaps over the 100 test load profiles. Each sample in the distribution corresponds to one load-profile instance: for that load profile, we first compute the enumeration-based benchmark worst-case value over the considered outage set, and then compare it with the value obtained from the topology selected by each NN- or PCNN-based method.
As shown in Fig.~\ref{fig:WLS_14}, PCNN produces optimality gaps tightly concentrated near zero, while direct NN maximization exhibits wider dispersion and larger high-gap cases.
The remaining larger-gap PCNN cases are relatively rare and are mainly associated with the AC relaxation used in the PCNN formulation. Although the relaxation is tight for most topologies, it can occasionally introduce enough discrepancy (see Fig.~\ref{fig:QC_relax}) in the relaxed lower-level objective to alter the ranking among near-worst topologies. Consequently, the PCNN may select a topology that is worst under the relaxed model but only near-worst when re-evaluated using the original AC-NLP benchmark. This relaxation-induced misranking helps explain the residual PCNN optimality gaps.\\

\noindent
\textbf{Performance of Area-Partitioned Surrogates}\\
We compare AWLS performance using the partitioned, block-sparse NN\_Multi against NN\_Single for both surrogate formulations. The partitioned architecture's sparsity reduces model complexity by dedicating each NN to electrically cohesive clusters. As Table~\ref{WLS_Follower_Leader} shows, on average, NN\_Multi yields significant gap reductions: direct-NN drops from 8.7\% to 7.6\%, while PCNN shows more substantial improvement from 5.8\% to 3.7\%. These gains confirm the partitioned approach better captures localized load-shed patterns. Computationally, on average, NN\_Multi reduces solve times from 0.93s to 0.89s (direct-NN) and 16.4s to 12.8s (PCNN), demonstrating that the localized design enhances both accuracy and efficiency for large-scale systems.

\begin{figure}[t!]
	\centering
	\subfloat[NN surrogate]{\includegraphics[scale=0.27]{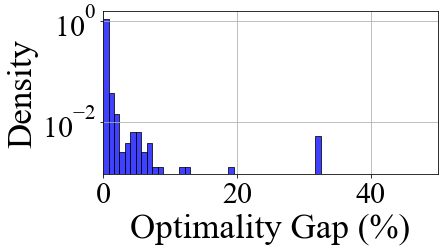}}
	\quad
	\subfloat[PCNN surrogate]{\includegraphics[scale=0.27]{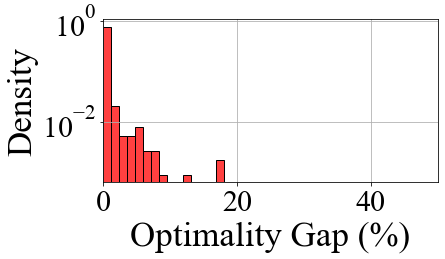}}
	\caption[]{\small Comparisons of the optimality gap of (a) NN surrogate and (b) PCNN surrogate with NN\_Multi for the 118-bus test systems.}\label{fig:WLS_118_mul}
\end{figure}

\subsection{Out-of-sample Performance for Practical Scenarios}
To demonstrate practical applicability, we evaluate our methods under realistic data where training examples cannot cover all adversarial topologies.
The proposed NN/PCNN formulation is not restricted to a pre-selected candidate line set. In principle, $\mathcal{X}(k)$ can be defined over any admissible set of outageable lines, including the full transmission-line set. However, enlarging this set substantially increases the combinatorial data-generation burden and the risk of surrogate distribution shift, especially for line outages that are poorly represented in training.
Unlike prior studies (such as \cite{aquino2023comparing,johnson2022scalable,liu2013optimal}) assuming complete data availability, real-world grid operations face combinatorial explosion in contingency scenarios. We test both direct-NN and PCNN surrogates with NN\_Multi-based sparse value-function approximation under two practically relevant data-constrained conditions: \\
(i) Unseen/Larger Damage Severities, reflecting the model's ability to handle severities beyond the training scenarios, and \\
(ii) Out-of-sample Line Sets, assessing the model’s performance when disruptions occur on lines not fully represented in the training dataset.

Both tests demonstrate that the proposed PCNN surrogate is more robust and accurate than simpler NN surrogates.

\subsubsection{Generalization to Unseen Damage Severities}
In the first practical scenario, we test the surrogate model's robustness to damage severities not seen in training. Training data are restricted to topologies with up to three simultaneous line outages; operational conditions may involve up to five line outages. To improve generalization, we first train on $\leqslant$3-outage topologies and then selectively augment the set with 10\% of higher-severity cases (four and five outages) drawn from the top 30\% worst-case impacts, explicitly exposing the model to critical damage patterns.

Table~\ref{WLS_application} reports that the NN surrogate yields an average optimality gap of 13.3\%, while the PCNN surrogate reduces it to 6.5\%; the maximum gap drops significantly from 39.4\% (NN) to 19.2\% (PCNN). Although this accuracy gain increases average solve time from 1.73 s (NN) to 14.5 s (PCNN), Fig.~\ref{fig:WLS_118_case1} shows consistently narrower gap distributions and improved accuracy for extreme scenarios. Leveraging the partitioned, block-sparse (NN\_Multi) architecture with cluster-specific inputs, the approach balances runtime and accuracy, making it highly suitable for operations that require rapid decisions and robustness to previously unseen, high-severity events.

\subsubsection{Generalization to Out-of-Sample Line Sets}

In the second practical scenario, we address the practical limitation where training data cover only a subset of critical lines due to resource constraints or incomplete historical records, creating generalization challenges when adversarial events occur on underrepresented lines. To mitigate this, we train on 15 critical lines augmented with 10\% of outage combinations from a broader 31-line set, providing limited exposure to diverse topologies while enhancing out-of-sample generalization. Testing uses the complete 31-line set to evaluate performance on topologies with lines not fully represented during training.

\begin{figure}[t!]
	\centering
	\subfloat[NN surrogate]{\includegraphics[scale=0.27]{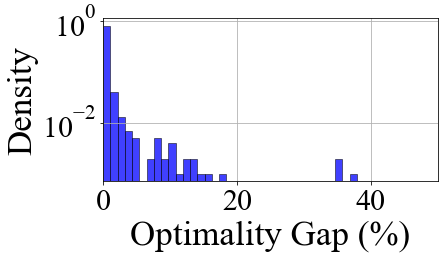}}
	\quad
	\subfloat[PCNN surrogate]{\includegraphics[scale=0.27]{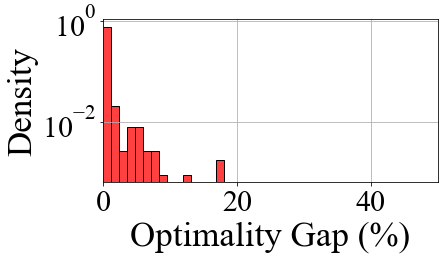}}
    \quad 
    \subfloat[NN surrogate]{\includegraphics[scale=0.27]{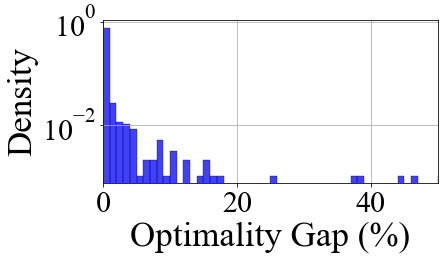}}
	\quad
	\subfloat[PCNN surrogate]{\includegraphics[scale=0.27]{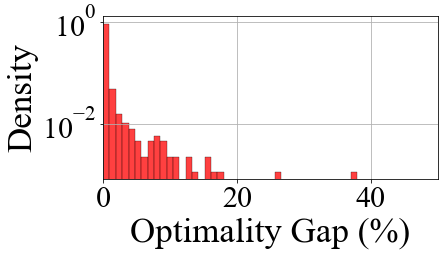}}
	\caption[]{\small Optimality gap comparisons on the 118-bus system under distribution shift: (a–b) unseen damage severities and (c–d) out-of-sample interdicted line sets; panels (a,c) show the NN surrogate, and panels (b,d) show the PCNN surrogate.}\label{fig:WLS_118_case1}
\end{figure}

Table~\ref{WLS_application} shows PCNN, when compared against the direct-NN surrogate, substantially improves out-of-sample accuracy: average optimality gap decreases from 17.5\% to 11.2\%, and maximum gap from 49.5\% to 37.4\%. Although compute time increases from 2.03s to 18.4s, Fig.~\ref{fig:WLS_118_case1} demonstrates markedly tighter gap distributions for PCNN, especially for severe contingencies. In summary, the sparsified, area-partitioned architecture (NN\_Multi) captures localized outage behavior sparsely represented in training, yielding robust generalization that suits large-scale contingency analysis where comprehensive training data are inherently limited.

\renewcommand{\arraystretch}{1.2}
\begin{table}[t!]
\caption{\small AWLS problem comparisons of two large dataset applications for both NN and PCNN surrogates on the 118-bus system.}
\begin{center}
\begin{tabular}{c|c|c|c|c|c}
\Xhline{3\arrayrulewidth}
  &  & \multicolumn{2}{c|}{\makecell[c]{Unseen damage\\severities}} & \multicolumn{2}{c}{\makecell[c]{Out-of-sample\\line sets}} \\ \cline{3-6}
  & & NN & PCNN & NN & PCNN \\ \hline
\multirow{3}{*}{{\makecell[c]{Optimality\\ Gap (\%)}}}  & min &  0\%    &  0\%     & 0\%    &   0\% \\
& avg &  13.3\%    &  \textbf{6.5\%}     & 17.5\%    &  \textbf{11.2\%} \\
& max &  39.4\%    &  19.2\%     & 49.5\%    &  37.4\% \\ \hline
\multirow{3}{*}{{\makecell[c]{Computation\\Time (s)}}} & min  & 1.57 s   &  12.7 s    & 1.89 s   &  17.5 s \\
& avg  & \textbf{1.73 s}   &  14.5 s    & \textbf{2.03 s}   &  18.4 s \\
& max  & 2.02 s   &  16.3 s    & 2.25 s   &  19.3 s \\
\Xhline{3\arrayrulewidth}
\end{tabular}\label{WLS_application}
\end{center}
\end{table}

\section{Conclusions} \label{sec:con}

We developed a novel neural network (NN)-based framework for efficiently solving adversarial worst-case load shedding (AWLS) problems, addressing the computational challenges of bilevel optimization with nonconvex AC power flow constraints. Our approach introduced two key innovations: an optimal value-function reformulation that embeds NN surrogates within physics-constrained power flow equations, and a scalable architecture that leverages spectral clustering to create partitioned, block-sparse NNs with linear parameter scaling. The resulting methodology transformed the bilevel AWLS problem into a tractable single-level mixed-integer convex program. Through extensive numerical validation on medium-scale IEEE systems, area-partitioned physics-constrained NN surrogate consistently outperformed naive NN surrogates in solution quality while maintaining practical computation times under one minute. Moreover, our architecture robustly generalized under realistic data limitations, including scenarios with higher damage severities and underrepresented transmission line outages in training data. \textit{To summarize, we validated sparse neural approximations as a scalable approach to practical bilevel adversarial problems in power grids.}

Future work will focus on (i) improving robustness to distributional shift across a wider range of contingencies and damage patterns, by developing adaptive sampling strategies to reduce the dependence on pre-screened candidate sets; (ii) extending the framework to time-coupled tasks such as dynamic restoration and carbon-aware operations; (iii) incorporating more operationally detailed load-shedding models, including power-factor-coupled active and reactive load shedding; and (iv) designing advanced NN architectures and formulations to obtain compact MILP models via binary variable reduction.

\bibliographystyle{IEEEtran}
\bibliography{Ref.bib}

\section*{Appendix} \label{appendix}
\noindent
Convex relaxations of the power flow in \eqref{eq:powerflow} allow for transforming the inherently non-convex formulations into more tractable convex surrogates. These surrogates make it possible to obtain both valid optimality bounds for global optimization algorithms and measures of suboptimality for local methods.
By using the lifted variables $w^x_{ij}:=x_{ij} V^2_i$, $w^c_{ij}:=x_{ij} V_i V_j \cos\theta_{ij}$, and $w^s_{ij}:=x_{ij} V_i V_j \sin\theta_{ij}$, the on/off switching power flow can be expressed in the linear form:

\begin{subequations}
\small
\begin{align}
    P_{ij} &= w^x_{ij} \big(\frac{ g_{ij}}{ a^2_{ij}}\!+\!g^{\mathrm {sh}}_i\big)\!-\! \frac{1}{ a_{ij}} \big( g_{ij}w^c_{ij} + b_{ij}w^s_{ij}\big),\\
    Q_{ij} &= - w^x_{ij} \big(\frac{ b_{ij}}{a^2_{ij}}\!+\!b^{\mathrm {sh}}_i\big) \! -\! \frac{1}{ a_{ij}} \big( g_{ij}w^s_{ij} - b_{ij}w^c_{ij}\big).
    \end{align}
\end{subequations}
The non-convexity is now encapsulated in the definitions of these new lifted variables. The quadratic convex relaxation model (QCRM) is then formed by defining the convex envelopes (i.e., relaxations) for $w^x_{ij}$, $w^c_{ij}$, and $w^s_{ij}$.

For the $w^x_{ij}$ term, the relaxation is built in two parts. First, the voltage-square term $w_i:=V^2_i$ is relaxed using its convex envelope:
\begin{align}
w_i & \geqslant V^2_i, \quad w_i \leqslant (\underline{V}_i+\bar{V}_i)V_i - \underline{V}_i \bar{V}_i,
\end{align}
where $\bar{V}_i$ and $\underline{V}_i$ are the bounds on $V_i$. Second, the bilinear on/off term $w^x_{ij} = x_{ij} w_i$ is relaxed using a McCormick formulation:
\begin{align}
w_i & - (1-x_{ij})\bar{V}^2_i \leqslant w^x_{ij} \leqslant w_i - (1-x_{ij})\underline{V}^2_i.
\end{align}

For the on/off trilinear terms $w^c_{ij}$ and $w^s_{ij}$, this QCRM uses a sequential McCormick relaxation. This approach decomposes the trilinear term into a series of bilinear products, which are then relaxed sequentially.
First, we define intermediate variables for the components: $w_{ij} := V_i V_j$, $c^x_{ij} := x_{ij} \cos\theta_{ij}$, and $s^x_{ij} := x_{ij} \sin\theta_{ij}$.
The on/off trigonometric terms ($c^x_{ij}, s^x_{ij}$) are relaxed using the disjunctive relaxation method from~\cite{bestuzheva2020convex}. For each $(i,j)$, let the phase-angle difference satisfy
$\underline{\theta}_{ij} \leqslant \theta_{ij} \leqslant \overline{\theta}_{ij}$ with
$-\tfrac{\pi}{2} \leqslant \underline{\theta}_{ij} \leqslant \overline{\theta}_{ij} \leqslant \tfrac{\pi}{2}$,
and define
\[
\bar{\theta}_{ij} := \max\{|\underline{\theta}_{ij}|,|\overline{\theta}_{ij}|\}.
\]
We also introduce a big-$M$ constant $M^\theta_{ij}$ such that $|\theta_{ij}| \leqslant M^\theta_{ij}$ when $x_{ij} = 0$.

The on/off cosine term $c^x_{ij} = x_{ij}\cos\theta_{ij}$ is relaxed using the curvature parameter $\kappa_{ij} := \frac{1 - \cos(\bar{\theta}_{ij})}{\bar{\theta}_{ij}^2},$ leading to the following quadratic constraints:
\begin{align*}
\small
\label{eq:onoff_cos}
    & x_{ij}\cos(\bar{\theta}_{ij}) \leqslant c^x_{ij} \leqslant x_{ij}, \\
    & c^x_{ij}
    \leqslant x_{ij}
      - \kappa_{ij}\,\theta_{ij}^2
      + (1-x_{ij})\,\kappa_{ij}\,(M^\theta_{ij})^2.
\end{align*}

The on/off sine term $s^x_{ij} = x_{ij}\sin\theta_{ij}$ is relaxed using a few constants,
\[
c_{ij} := \cos\!\Bigl(\tfrac{\bar{\theta}_{ij}}{2}\Bigr), \quad
s_{ij} := \sin\!\Bigl(\tfrac{\bar{\theta}_{ij}}{2}\Bigr), \quad 
\alpha_{ij} := s_{ij} - c_{ij}\tfrac{\bar{\theta}_{ij}}{2}, 
\]
\[
\beta_{ij} := s_{ij} + c_{ij}\tfrac{\bar{\theta}_{ij}}{2}, \quad
\gamma_{ij} := \alpha_{ij} + \sin(\bar{\theta}_{ij}), \quad
M^{c}_{ij} := c_{ij} M^\theta_{ij},
\]
and, the corresponding linear relaxation is
\begin{align*}
\small
\label{eq:onoff_sin}
    & x_{ij} \sin(\underline{\theta}_{ij}) \leqslant s^x_{ij} \leqslant x_{ij} \sin(\overline{\theta}_{ij}), \\
    & \sigma\, s^x_{ij} - \sigma\, c_{ij}\,\theta_{ij}
    \leqslant x_{ij}\alpha_{ij} + (1-x_{ij}) M^{c}_{ij},
    &&\forall \sigma \in \{-1,1\}, \\
    & \sigma\, s^x_{ij}
    \leqslant x_{ij}\beta_{ij},
    &&\forall \sigma \in \{-1,1\}, \\
    & \sigma\, c_{ij}\,\theta_{ij}
    \leqslant x_{ij}\gamma_{ij} + (1-x_{ij}) M^{c}_{ij},
    &&\forall \sigma \in \{-1,1\}.
\end{align*}

The voltage product $w_{ij}$ is relaxed using the standard 4-inequality McCormick envelope $\langle V_i V_j \rangle^M$, defined as:
\begin{subequations}
\begin{align}
    \langle vw \rangle^M &\equiv (\tilde{vw}, v, w) \in \mathbb{R}^3 \nonumber\\
    &\text{s.t.}
    \begin{cases}
        \tilde{vw} \geqslant v^l w + w^l v - v^l w^l, \nonumber\\
        \tilde{vw} \geqslant v^u w + w^u v - v^u w^u, \nonumber\\
        \tilde{vw} \leqslant v^l w + w^u v - v^l w^u, \nonumber\\
        \tilde{vw} \leqslant v^u w + w^l v - v^u w^l. \nonumber
    \end{cases}
\end{align}
\end{subequations}
Finally, the full trilinear terms are relaxed by applying another McCormick relaxation to the product of the intermediate relaxed variables:
\begin{align}
w^c_{ij}=\langle {w}_{ij} {c}^x_{ij} \rangle^M ~\text{and}~ w^s_{ij}=\langle {w}_{ij} {s}^x_{ij} \rangle^M.
\end{align}
This relaxation is computationally efficient and provides a strong bound. For a detailed discussion, see~\cite{bestuzheva2020convex,guo2025tightening}.

\ifCLASSOPTIONcaptionsoff
\newpage
\fi

\end{document}